\documentclass[prl,twocolumn,aps,superscriptaddress,notitlepage,floatfix,10pt]{revtex4-2}
\usepackage{amsthm,amsmath,amssymb,amsfonts}
\usepackage{booktabs} 
\usepackage{multirow}
\usepackage[utf8]{inputenc}
\usepackage[american]{babel}
\usepackage{graphicx,xcolor,bbold,titlesec}
\usepackage{MnSymbol}
\usepackage{braket}
\usepackage{comment}
\usepackage{scalerel}
\usepackage{lipsum}
\usepackage{xcolor}
\usepackage[colorlinks,bookmarks=false,citecolor=blue,linkcolor=blue,urlcolor=blue]{hyperref}
\usepackage{simplewick}
\usepackage{wasysym}
\usepackage[toc]{appendix}
\usepackage{graphics}
\usepackage{multirow}
\usepackage{algpseudocode}
\usepackage{dsfont}
\usepackage{mdframed}
\usepackage{tikz,ifthen}
\usepackage{tikz-network}
\usepackage{orcidlink}

\usetikzlibrary{patterns,decorations.pathreplacing,calligraphy}
\usetikzlibrary{shapes,arrows.meta,decorations.pathmorphing}

\renewcommand{\section}[1]{\textbf{\emph{#1}}.\!}
\renewcommand{\subsection}[1]{{\emph{#1}}.---}

\newcommand{\bbra}[1]{\llangle #1 |}
\newcommand{\kket}[1]{| #1 \rrangle}
\newcommand{\bbrakket}[2]{\llangle #1 | #2 \rrangle}

\begin{document}

\title{Anticoncentration and State Design of Doped Real Clifford Circuits and Tensor Networks}

\author{Beatrice Magni~\orcidlink{0009-0009-8577-0525}}
\email{bmagni@uni-koeln.de}
\affiliation{Institut für Theoretische Physik, Universität zu Köln, Zülpicher Strasse 77, 50937 Köln, Germany}

\author{Markus Heinrich~\orcidlink{0000-0002-1334-7885}}
\affiliation{Institut für Theoretische Physik, Universität zu Köln, Zülpicher Strasse 77, 50937 Köln, Germany}

\author{Lorenzo Leone~\orcidlink{0000-0002-0334-7419}}
\affiliation{Dipartimento di Ingegneria Industriale, Università degli Studi di Salerno,
Via Giovanni Paolo II, 132, 84084 Fisciano (SA), Italy}
\affiliation{Dahlem Center for Complex Quantum Systems,
Freie Universität Berlin, 14195 Berlin, Germany}

\author{Xhek Turkeshi~\orcidlink{0000-0003-1093-3771}}
\email{xturkesh@uni-koeln.de}
\affiliation{Institut für Theoretische Physik, Universität zu Köln, Zülpicher Strasse 77, 50937 Köln, Germany}

\date{\today}

\begin{abstract}
We investigate the statistical properties of orthogonal, or real, Clifford circuits doped with magic and imaginary resources. By developing the Weingarten calculus for the real Clifford group, we derive the exact overlap distribution of real stabilizer states, identifying a new universality class: the orthogonal Clifford Porter-Thomas distribution. We prove that local real architectures recover this global statistic in logarithmic depth. Furthermore, we uncover a sharp hierarchy in resource requirements: while retrieving Haar statistics necessitates a polylogarithmic amount of magic states, recovering the full unitary Clifford statistics requires only a single phase gate.
\end{abstract}

\maketitle

\section{Introduction} 
Clifford circuits and magic states are central ingredients to fault-tolerant quantum computing~\cite{gottesman1997stabilizer,gottesman1998theory,bravyi2005universal}. 
Indeed, Clifford circuits drive quantum error correction and admit efficient fault-tolerant implementations.
However, Clifford dynamics alone are efficiently classically simulatable~\cite{aaronson2004improved,Bravyi2016,Bravyi2019simulationofquantum,sierant2023entanglement}. 
Only the additional supply of magic states can surpass efficient simulation schemes and enable universal quantum computation.
Clifford circuits alone, or ``doped'' with a controlled number of magic resources, have proven an invaluable tool in various fields, from the characterization of out-of-equilibrium systems~\cite{Leone2021quantumchaosis, magni2025quantumcomplexitychaosmanyqudit, lami2024unveiling,aditya2025mpembaeffectsquantumcomplexity,tirrito2025universalspreadingnonstabilizernessquantum,Turkeshi2025,tirrito2025anticoncentration,falcao2025nonstabilizerness,odavic2025stabilizerentropynonintegrablequantum, bejan2024dynamical, hou2025stabilizerentanglementenhancesmagic,varikuti2025impactcliffordoperationsnonstabilizing, szombathy2025spectralpropertiesversusmagic, szombathy2025independentstabilizerrenyientropy, niroula2023phase,Turkeshi2024, zhou2020single}, to quantum applications~\cite{ yoganathan2019quantum, leone2025noncliffordcostrandomunitaries,lóio2025quantumstatedesignsmagic, liu2024classicalsimulabilitycliffordtcircuits, haug2024efficient, haferkamp2022, Mao_2025, haug2024probingquantumcomplexityuniversal, Chia2024efficientlearningof, hinsche2023one, fux2024disentanglingunitarydynamicsclassically, harper2025gcampsscalableclassicalsimulator,nakhl2024stabilizertensornetworksmagic}, even touching questions of quantum gravity and black-hole physics~\cite{leone2022retrieving, leone2024learning, oliviero2024unscrambling, haferkamp2022linear, Yoshida_2019}.

Crucial to a wide range of applications is the ability of doped Clifford circuits to generate random quantum states.
Two fundamental ways of quantifying this are anticoncentration and state designs. 
The former implies that the circuit spreads probability amplitudes broadly across the Hilbert space rather than localizing on a few basis states~\cite{Bouland2018on, boixo2018characterizing, hangleiter_anticoncentration_2018, dalzell2022random, heinrich2025anticoncentrationalmostneed}. 
The latter ensures that the output ensemble reproduces the lower-order moments of the Haar measure~\cite{nakata2025computational,lami2025anticoncentration, Christopoulos_2025}.
Previous results on doped Clifford circuits established that a logarithmic amount of magic states suffices to mimic these statistical properties of Haar-random states~\cite{magni2025anticoncentrationcliffordcircuitsbeyond}. 
However, the statistical properties of further restricted ensembles remain largely unexplored. 
Here, we address this gap by characterizing the anticoncentration and state design properties of \emph{real, or orthogonal, Clifford} circuits and tensor networks, both in the pure regime and when doped with controlled amounts of magic~\cite{liu2022manybody, chitambar2019quantum} or imaginary resources~\cite{wu_immresource2021, deneris_analyzing_2025, shi_deimginarity_2025, shi_erasing_2025} injected into them. We address three key questions: \textit{(i) What characterizes the output distribution and state design of random real Clifford states? (ii) Under what conditions do local tensor networks or circuits recover these global statistics? (iii) How does doping with different resources affect the anticoncentration and state design?}

To address these questions, we exploit the Weingarten calculus for the real Clifford group.
Combining this with the stabilizer tableau formalism, we derive rigorous results for random tensor networks and local circuits, identifying a distinct universality class: the \textit{orthogonal Clifford Porter-Thomas} (OCPT) distribution. We show that Random Matrix Product States (RMPS) and shallow local circuits saturate to this distribution with polynomial bond dimension and logarithmic depth, respectively. Furthermore, we characterize the impact of resource doping: while polylogarithmic real or complex magic is required to approach Haar statistics, a single imaginary state suffices to recover unitary Clifford statistics strictly. 
These findings establish a sharp separation of scales between magic and imaginarity, quantifying the distinct costs required to access orthogonal and unitary state designs.

\begin{table*}[t!]
    \centering
    \setlength{\tabcolsep}{5pt}
    \begin{tabular}{|l|c|l|c|l|c|}
        \hline
        \textbf{Dynamical Ensemble} & \textbf{Complexity Cost} & \textbf{Doping Resource} & \textbf{Amount} & \textbf{State-design} & \textbf{Ref.} \\
        \hline
        unitary Haar ($\mathcal{U}$) & 
        \multirow{5}{*}{\shortstack{Circuit [$\mathrm{depth}\sim \log N$] \\ RMPS  [$\chi \sim \text{poly}(N)$]}} & 
        \multicolumn{1}{c|}{---} & --- & $\mathbb{C}$Haar & \cite{schuster2025randomunitariesextremelylow,lami2025anticoncentration} \\ \cline{1-1} \cline{3-6}
        \multirow{2}{*}{orthogonal Haar ($\mathcal{O}$)} & & 
        \multicolumn{1}{c|}{$\varnothing$} & --- & $\mathbb{R}$Haar & {\cite{magni2025anticoncentrationcliffordcircuitsbeyond,zhang2025designsmagicaugmentedcliffordcircuits}} \\ \cline{3-6}
         & & Imaginary States $|\pm i\rangle$ & $O(1)$ & $\mathbb{C}$Haar & ~\cite{haug_pseudorandom_2024,SM}\\
         \cline{1-1}\cline{3-6}
        \multirow{2}{*}{unitary Clifford ($\mathcal{C}$)} & & 
        \multicolumn{1}{c|}{$\varnothing$} & --- & $\mathbb{C}$Stab & \multirow{2}{*}{\cite{magni2025anticoncentrationcliffordcircuitsbeyond,zhang2025designsmagicaugmentedcliffordcircuits}} \\ \cline{3-5}
         & & Real/Complex Magic States $|H\rangle$ & $O(\log N)$ & $\mathbb{C}$Haar & \\
        \hline\hline
        \multirow{4}{*}{\shortstack{\textbf{orthogonal Clifford} ($\mathcal{R}$)\\ \textbf{(This work)}}} & 
        \multirow{4}{*}{\shortstack{Circuit [$\mathrm{depth}\sim \log N$] \\ RMPS  [$\chi \sim \text{poly}(N)$]}} & 
        \multicolumn{1}{c|}{$\varnothing$} & --- & $\mathbb{R}$Stab & Eq.~\eqref{eq:scalingCRMPS} \\ \cline{3-6}
         & & Real Magic $|H\rangle$ & $O(\log N)$ & $\mathbb{R}$Haar & Eq.~\eqref{eq:dopedIPROrtho} \\ \cline{3-6}
         & & Complex Magic $|T\rangle$ & $O(\log N)$ & $\mathbb{C}$Haar & Eq.~\eqref{eq:dopedIPRunitary} \\ \cline{3-6}
         & & Imaginary state $|\pm i\rangle$ & $O(1)$ & $\mathbb{C}$Stab & Eq.~\eqref{eq:dopedIPRClifford} \\
        \hline
    \end{tabular}
    \caption{\footnotesize Comparison of state designs. Note the vertical and horizontal separators emphasizing the resource transitions.}
    \label{tab:results}
\end{table*}

\section{Methods}
We consider an $N$-qubit system with Hilbert space dimension $d \equiv 2^N$, initialized in the state $|\psi_0\rangle$. 
The system evolves via a depth-$t$ quantum circuit, yielding $|\psi\rangle = \mathbf{C}_t |\psi_0\rangle$.
The circuit is decomposed into layers as $\mathbf{C}_t = \prod_{s=1}^t \left( \bigotimes_{\lambda \in \Lambda_s} C_\lambda \right)$, where $\Lambda_s$ specifies the set of interaction supports active in layer $s$. 
Each local gate $C_\lambda$ is a random orthogonal Clifford, i.e. $C_\lambda \in \mathcal{R}_n$ s.t. $C_\lambda C_\lambda^T=\mathbb{1}$, and acts on the contiguous qubits in $\lambda$~\footnote{Explicitly, $C_\lambda \equiv C_{\lambda_1,\ldots,\lambda_n}
= \mathbb{I}^{\otimes(\lambda_1-1)} \otimes C \otimes \mathbb{I}^{\otimes(N-\lambda_n)}$,
where $n=|\lambda|$ depends on the geometry under consideration and $C\in\mathcal{R}_{n}$. For simplicity, we use the shorthand
$C_\lambda\in\mathcal{R}_{n}$.}.

We defer the detailed definition of the architectures $\mathfrak{A}=\{\Lambda_s\}_{s=1}^t$ to the next section. Here, we introduce the central quantities of interest.
To probe anticoncentration, we study the probability distribution of the scaled output probabilities $w = d |\langle \mathbf{y}|\psi\rangle|^2$
\begin{equation}
    P(w) = \mathbb{E}_{\mathbf{C}_t}\left[ \frac{1}{d} \sum_{\mathbf{y} \in \mathcal{B}} \delta\left(w - d |\langle \mathbf{y}|\psi\rangle|^2\right) \right],
    \label{eq:overlapdistrib}
\end{equation}
where $\mathcal{B} = \{ |\mathbf{y}\rangle \}_{\mathbf{y}=0}^{d-1}$ is the computational basis and the circuit average $\mathbb{E}_{\mathbf{C}_t}[\bullet]$ is over each independent gate. 
The distribution~\eqref{eq:overlapdistrib} is equivalently captured by the Inverse Participation Ratios (IPRs)~\cite{luitz2014participation, mace2019multifractal, liu2024quantum, sierant2022mipt,sierant2022universal}, which are proportional to its moments
\begin{equation}
    I_k \equiv \mathbb{E}_{\mathbf{C}_t} \sum_{\mathbf{y}\in \mathcal{B}} |\langle \mathbf{y}|\psi\rangle|^{2k} = d^{1-k} \int w^k P(w) dw\,.
    \label{eq:ipr}
\end{equation}
Complementarily, we assess state $k$-design properties via the frame potential~\cite{deluca2024universalityclassespurificationnonunitary},
\begin{equation}
    \mathcal{F}_k \equiv \mathbb{E}_{\mathbf{C}_t, \mathbf{C}_t'} [|\langle \psi' | \psi \rangle|^{2k}]\,,
    \label{eq:framepot}
\end{equation}
which measures the overlap between states $|\psi\rangle$ and $|\psi'\rangle$ generated by two random circuits $\mathbf{C}_t$ and $\mathbf{C}_t'$.

All aforementioned quantities involve statistical averages over the ensemble of real Clifford gates. To evaluate them, we employ the \textit{Weingarten calculus}, formulated in the vectorized representation $A \mapsto \kket{A}$ equipped with the inner product $\bbrakket{A}{B} =\mathrm{tr}(A^\dagger B)$. In this formalism, we have $|U A U^\dagger \rrangle= (U\otimes U^*)\kket{A}$~\cite{Jamiokowski1972, mele2024introduction}.
The key identity of the calculus expresses the $k$-th moment of $\mathcal{R}_n$ as
\begin{equation}
    \mathbb{E}_{U \in \mathcal{R}_n}[(U \otimes U^\ast)^{\otimes k}] = \sum_{\sigma,\pi \in \Xi_k} \mathrm{Wg}_{\sigma,\pi}(2^n) \kket{\sigma}\bbra{\pi}\,,
    \label{eq:shur_weil}
\end{equation}
where the sum runs over elements of a suitable basis $\Xi_k$ of the commutant algebra $\mathrm{Comm}_{k}(\mathcal{R}_n)= \{ A \mid [A, U^{\otimes k}] = 0, \forall\, U \in \mathcal{R}_n \}$ of the orthogonal Clifford group.
The coefficients $\mathrm{Wg}_{\sigma,\pi}$ constitute the Weingarten matrix, defined as the (pseudo) inverse of the Gram matrix of the basis overlaps, $G_{\sigma,\pi} = \bbrakket{\sigma}{\pi}$.
We briefly state relevant properties in the following and refer to the Supplemental Material~\cite{SM} for the explicit commutant construction and proofs.

First, defining the \textit{real stabilizer purity} associated with any basis element $\sigma \in \Xi_k$ as $\zeta_\sigma(|\psi\rangle) \equiv \mathrm{tr}[\sigma (|\psi\rangle\langle\psi|)^{\otimes k}]$, we show that (i) $0\le \zeta_\sigma\le 1$ for generic states and $\zeta_\sigma=1$ for real stabilizer states, $\mathbb{R}\mathrm{Stab}_N \equiv \{\ket{\psi} = C\ket{0}^{\otimes N}\,|\, C\in \mathcal{R}_N\}$; (ii) it factorizes across tensor products, satisfying $\zeta_\sigma(|\psi\rangle \otimes |\phi\rangle) = \zeta_\sigma(|\psi\rangle)\zeta_\sigma(|\phi\rangle)$. 
Second, the dimension of the real Clifford commutant for $N \ge k$ is given by the $q$-Pochhammer symbol, $|\Xi_k| = (-2; 2)_k$, where $(a; q)_n = \prod_{j=0}^{n-1} (1 - a q^j)$~\cite{Kac2002}.
Finally, the Weingarten and Gram matrices satisfy the summation identity
\begin{equation}
    \sum_{\pi \in \Xi_k} \mathrm{Wg}_{\pi,\sigma}(d) = \frac{d^{-k}}{(-2 d^{-1}; 2)_k} = \left[ \sum_{\pi \in \Xi_k} G_{\pi,\sigma}(d) \right]^{-1} \;.
    \label{eq:Wsum}
\end{equation}

\section{Results} 
We start deriving the overlap distribution of real random stabilizer states obtained by acting with a global unitary $C\in\mathcal{R}_N$ on the reference state $\ket{\boldsymbol{0}}=\ket{0}^{\otimes N}$.
Since a real Clifford maps $\boldsymbol{y}$ to $\boldsymbol{0}$, Eq.~\eqref{eq:overlapdistrib} reduces to $P(w)=\mathbb{E}_{C\in\mathcal{R}}[\delta(w-d|\langle\boldsymbol{0}|C|\boldsymbol{0}\rangle|^2)]$. 

Any real stabilizer state $|\psi\rangle \in \mathbb{R}\mathrm{Stab}_N$ can be expressed as $|\psi\rangle = 2^{-g/2} \sum_{i=0}^{2^g-1} (-1)^{\varphi_i} |\mathbf{x}_i\rangle$, where $0 \le g \le N$ and $\varphi_i \in \{0,1\}$. 
Consequently, the overlap distribution is uniformly supported on the set $\mathcal{B}_\psi = \{ \mathbf{x}_i \}$ of size $2^g$, given by $|\langle \mathbf{y} | \psi\rangle|^2 = 2^{-g} \mathbb{1}_{\mathcal{B}_\psi}(\mathbf{y})$ with $\mathbb{1}_A$ the characteristic function of $A$. 
In the stabilizer tableau formalism~\cite{SM,nielsen2010quantum}, the parameter $g$ corresponds to the participation entropy, determined by the rank of the $\mathbf{X}$-block $g = \mathrm{rk}_{\mathbb{Z}_2}(\mathbf{X})$~\cite{sierant2022universal,turkeshi2023measuring}.

To derive the ensemble properties, we enumerate states sharing a fixed entropy $g$.
For complex stabilizer states, this count is $\aleph_g = 2^N 2^{g(g+1)/2} \binom{N}{g}_2$~\cite{magni2025anticoncentrationcliffordcircuitsbeyond}, where $\binom{N}{g}_2$ denotes the Gaussian binomial coefficient.
Restricting to the real domain $\mathbb{R}\mathrm{Stab}_N$ requires, in the tableau formalism, an even number of $Y=iZ X$ operators per generator, which reduces the count by a factor of $2^g$. 
This yields the count for real stabilizer states
\begin{equation}
    \aleph_g^{\mathrm{real}} = \frac{\aleph_g}{2^g} = 2^N 2^{g(g-1)/2} \binom{N}{g}_2.
\end{equation}
Summing over all possible entropies recovers the total cardinality $|\mathbb{R}\mathrm{Stab}_N| = \sum_{g=0}^N \aleph_g^{\mathrm{real}} = 2^N (-1; 2)_N$.
Normalizing $\aleph_g^{\mathrm{real}}$ by this total count yields the distribution 
\begin{equation}
    \mathcal{P}_{\mathcal{R}}(g) = \binom{N}{g}_2 \frac{2^{g(g-1)/2}}{(-1; 2)_N}.
    \label{eq:HaarCdistribution}
\end{equation}
In the thermodynamic limit $N\to\infty$, introducing $n=N-g$ yields the limiting distribution
\begin{equation}
\mathcal{P}_{\mathcal{R}}(n)
=\frac{1}{(-1;2^{-1})_\infty}
\frac{2^{-n(n+1)/2}2^n}{(2^{-1};2^{-1})_n},
\label{eq:Orthocliff}
\end{equation}
which we term the \emph{orthogonal Clifford Porter--Thomas} (OCPT) distribution.
It defines a universality class distinct from both the unitary Clifford and orthogonal Haar ensembles.

\begin{figure*}[t!]
    \centering
    \includegraphics[width=\linewidth]{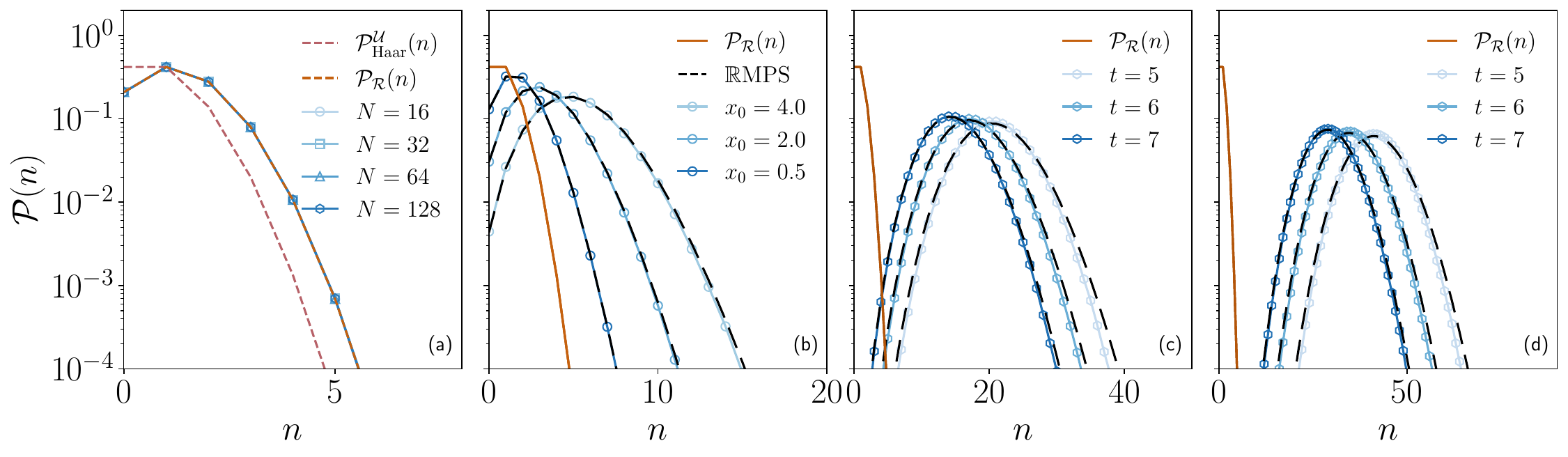}
    \caption{ Numerically sampled overlap distribution for different settings. Panel (a) presents the sampling of real random stabilizer states which agree with the distribution in Eq.~\eqref{eq:Orthocliff} (orange dashed line), compared to the unitary Clifford Porter-Thomas distribution (red dashed line). Panel (b) samples from $\mathbb{R}$MPS comparing the data with the distribution in Eq.~\eqref{eq:RMPSdistr} (black dashed line) for different $x_0 = N/\chi$ and the limiting distribution for real random stabilizer states. Panels (c) and (d) show the data sampled from a random brickwork circuit at different depths $t$ for $N = 128$ and $N=256$. Notably, the black dashed lines are obtained from the distribution~\eqref{eq:RMPSdistr} with the scaling variable $x$ as a single fitting parameter, while the orange line represents the OCPT.}
    \label{fig:distribution}
\end{figure*}

Using Eq.~\eqref{eq:HaarCdistribution} and the $q$-binomial theorem, the average inverse participation ratios read
\begin{equation}
I_k^{\mathcal{R}}
=\sum_{g=0}^N 2^{(1-k)g}\mathcal{P}_{\mathcal{R}}(g)
=
\frac{2^{(1-k)N}(-2;2)_{k-1}}{\left(-2^{-N+1};2\right)_{k-1}} .
\label{eq:IPRHaarCl}
\end{equation}
For $k\le3$, this exactly reproduces the moments of the orthogonal ensemble~\cite{sauliere2025universalityanticoncentrationchaoticquantum}, as expected by the 3-design property~\cite{webb2016cliffordgroupformsunitary,gross2021schurweyl}.
Finally, by Clifford invariance, the frame potential follows as $\mathcal{F}_k^{\mathcal{R}}
=\mathbb{E}_{C\sim\mathcal{R}}[|\bra{\mathbf{0}}C\ket{\mathbf{0}}|^{2k}]
=d^{-1}I_k^{\mathcal{R}}$. 
This relation shows that the state-design complexity $\mathcal{F}_k^{\mathcal{R}}$ is directly controlled by the anticoncentration moments $I_k^{\mathcal{R}}$.
As we show below, the same structure persists across different circuit architectures in the scaling limit.

We now describe the anticoncentration and state design properties of random matrix product states and quantum circuits built from real Clifford gates starting from the state $|\boldsymbol{0}\rangle$. 
First, we specialize the general framework to real random matrix product states ($\mathbb{R}$MPS). This ensemble is defined by a sequential "staircase" architecture, where the global unitary decomposes as $\mathbf{C}_{N-r} = \prod_{s=1}^{N-r} C_{\lambda_s}$. 
Here, each local gate $C_{\lambda_s}$ is drawn uniformly from the real Clifford group acting on the contiguous block of $r+1$ qubits $\lambda_s = \{s, \dots, s+r\}$. 
Crucially, the overlap of $r$ qubits between consecutive gates acts as a memory, enforcing an entanglement structure equivalent to an MPS with bond dimension $\chi = 2^r$~\cite{garnerone2010typicality, haferkamp2021emergent}. (Up to Clifford invariance, this ensemble can be recast in standard tensor network language, reshaping the matrices~\cite{lami2025anticoncentration}). 
Using local real Clifford invariance, Eq.~\eqref{eq:ipr} is recast as
\begin{equation}
    I_k^{\mathbb{R}\mathrm{MPS}}= d \bbra{\boldsymbol{0},\boldsymbol{0}}^{\otimes k}\mathbb{E}[(\mathbf{C}_{N-r} \otimes \mathbf{C}_{N-r}^\ast)^{\otimes k}] \kket{\boldsymbol{0},\boldsymbol{0}}^{\otimes k}\,.
\end{equation}
Exploiting the sequential architecture and Eq.~\eqref{eq:shur_weil}, we have 
\begin{equation}
\begin{split}
   & I_k^{\mathbb{R}\mathrm{MPS}} =\! \sum_{\{\pi_j,\sigma_j\}} \prod_{j=1}^{N-r-1} \left[ \mathrm{Wg}_{\pi_j,\sigma_j}(2\chi)G_{\sigma_j,\pi_{j+1}}(\chi) \right] \\
    &\quad \ \times \mathrm{Wg}_{\pi_{N-r},\sigma_{N-r}}(2\chi) \llangle \mathbf{0}|^{\otimes 2k} \!\cdot\! | \pi_1 \rrangle \llangle \sigma_{N-r} |\!\cdot\! | \mathbf{0} \rrangle^{\otimes 2k}
\end{split}\;.\label{eq:ziobelloo}
\end{equation}
$\ket{\mathbf{0}}$ is a real stabilizer state and the real stabilizer purity is $\llangle \pi |\!\cdot\! | \mathbf{0},\mathbf{0} \rrangle^{\otimes k}=1$ for all $\pi\in \Xi_k$.
We can then use the summation rule~\eqref{eq:Wsum} recursively, yielding the exact expression
\begin{equation}
\begin{split}
    I_k^{\mathbb{R}\mathrm{MPS}} 
    &= 2^{(1-k)N}(-2;2)_{k-1}\frac{\left[(-2^{-r+1};2)_{k-1}\right]^{N-r-1}}{\left[(-2^{-r};2)_{k-1}\right]^{N-r}}\;.
    \label{eq:IPRrandomMPS}
\end{split}
\end{equation}
In the scaling limit $N, \chi \to \infty$ with $x_0 = N/\chi$ fixed, an asymptotic expansion gives the compact form
\begin{equation}
    I_k^{\mathbb{R}\mathrm{MPS}} \simeq I_k^{\mathcal{R}} \exp\left[ \frac{2^k - 2}{2} x \right] \;,
    \label{eq:scalingCRMPS}
\end{equation}
where the effective scaling parameter $x \approx x_0 [1 - N^{-1} \log_2(N/x_0)]$ incorporates the leading logarithmic finite-size corrections. 
Rescaling the moments by $d^k$, we interpret Eq.~\eqref{eq:scalingCRMPS} as describing a composite random variable $n = n_0 + p$. Here, $n_0$ is drawn from the OCPT distribution, while $p$ follows an independent Poisson distribution $\mathcal{P}(p) = \lambda^p e^{-\lambda}/p!$ with rate $\lambda = x/2$.
Consequently, the probability distribution for the $\mathbb{R}\mathrm{MPS}$ ensemble is given by the convolution
\begin{equation}
    \mathcal{P}(n) = \frac{e^{- x} (2x)^n}{2^n (-1; 1/2)_{\infty}} \sum _{p=0}^n \frac{x^{-p} \, 2^{-p(p-1)/2}}{(n-p)! \, (1/2; 1/2)_{p}}
    \;.\label{eq:RMPSdistr}
\end{equation}

We extend these considerations to \textit{glued circuits}, a class of shallow architectures investigated in Refs.~\cite{schuster2025randomunitariesextremelylow,grevink2025glueshortdepthdesignsunitary,magni2025anticoncentrationcliffordcircuitsbeyond}.
Physically, these circuits partition the system into patches of $r$ qubits and apply gates acting on adjacent pairs of patches, covering $2r$ qubits per gate.
Formally, the evolution is given by $\mathbf{C}_{2} = \left(\prod_{j=0}^{{N}/{2r}-2} C_{\lambda_{2j+1} \cup \lambda_{2j+2}}\right) \left( \prod_{j=0}^{{N}/{2r}-1} C_{\lambda_{2j} \cup \lambda_{2j+1}}\right)$, 
where $\lambda_k$ denotes the patch $\{1+rk,\dots, (k+1)r\}$.
By construction, this architecture generates the same overlap distribution as the $\mathbb{R}\mathrm{MPS}$ with both physical and bond dimension $\chi=2^r$, cf.~Ref.~\cite{magni2025anticoncentrationcliffordcircuitsbeyond}. 
Consequently, the evaluation of the moments parallels the derivation above: the circuit average factorizes over the independent gates, yielding
\begin{equation}
     I_k^{\mathbb{R}\mathrm{GC}}  
    = I_k^{\mathcal{R}}(-2^{-N+1};2)_{k-1}\frac{\left[(-2^{-r+1};2)_{k-1}\right]^{\frac{N-2r}{r}}}{\left[(-2^{-2r+1};d)_{k-1}\right]^{\frac{N-r}{r}}}\,.
\end{equation}
In the scaling limit $N, \chi \to \infty$ with the ratio $x_0 = 2N/(\chi \log_2 \chi)$ held constant, we obtain the asymptotic behavior
\begin{equation}
    I_k^{\mathbb{R}\mathrm{GC}} = I_k^\mathcal{R} \exp \left[\frac{2^k-2}{2} x \right]\;,
\end{equation}
where the effective scaling parameter is $x \simeq x_0$ at leading order. 
In particular, the same probability distribution $\mathcal{P}(n)$ holds as for the $\mathbb{R}$MPS case~\eqref{eq:RMPSdistr}.
Crucially, this implies that $r \propto \log N$ suffices to suppress the exponential grow and drive the system toward the global random statistics.
Since a random Clifford unitary on $2r$ qubits can be implemented by a local circuit with depth linear in $r$, we conclude that local architectures achieve complete anticoncentration at a logarithmic depth $t \sim O(\log N)$.

We corroborate our analytical findings via stabilizer tableau simulations~\cite{aaronson2004improved}, generating random real Clifford unitaries according to the algorithm of Ref.~\cite{Hashagen2018}.
Figure~\ref{fig:distribution} summarizes the results.
Panel (a) demonstrates the convergence of the numerically sampled distribution of $n$ to the global OCPT prediction, Eq.~\eqref{eq:Orthocliff}, highlighting excellent agreement.
Similarly, for the $\mathbb{R}\mathrm{MPS}$ architecture, the simulations perfectly match the analytical convolution formula~\eqref{eq:RMPSdistr} across different values of the scaling ratio $x_0$.
Finally, we perform a stringent test on local brickwork circuits, defined by a staggered arrangement of nearest-neighbor gates, $\mathbf{C}_t = \prod_{s=1}^t \bigotimes_{j} C_{2j+(s\bmod 2), \, 2j+1+(s\bmod 2)}$.
By treating the variable $x$ in Eq.~\eqref{eq:RMPSdistr} as a single fitting parameter, we find that the theoretical distribution accurately captures the finite-size numerics, cf. Fig.~\ref{fig:distribution}(c,d).

Finally, we address the frame potential. For independent shallow circuits $\mathbf{C}_t$ and $\mathbf{C}'_t$, the overlap reduces to the return probability of a combined circuit of depth $2t$. Exploiting ensemble invariance, we find $\mathcal{F}_k = \mathbb{E}_{\mathbf{C}_{2t}}[|\langle \mathbf{0} | \mathbf{C}_{2t} | \mathbf{0} \rangle|^{2k}] \propto I_k$.
This linear relationship implies that a logarithmic circuit depth, $O(\log N)$, which is sufficient to establish anticoncentration, also suffices to drive the frame potential to its real Clifford value.
For the $\mathbb{R}\mathrm{MPS}$ ensemble, we analyze the frame potential in the large-$\chi$ limit using the asymptotic expansion of the orthogonal Weingarten calculus~\cite{gross2021schurweyl,bittel2025completetheorycliffordcommutant,magni2025anticoncentrationcliffordcircuitsbeyond}.
A lengthy but otherwise straightforward computation, closely paralleling the strategy employed in the unitary case~\cite{lami2025anticoncentration,dowling2025freeindependenceunitarydesign}, yields $\mathcal{F}_k^{\mathbb{R}\mathrm{MPS}} \simeq \mathcal{F}_k^{\mathcal{R}} \exp\!\left[(2^{k-1}-1)\, x \right]$, where the effective scaling parameter is renormalized to $x \propto N/\chi^2$~\footnote{A complete statistical field-theoretic treatment of Clifford circuits will be presented in a forthcoming work~\cite{magniToAp}.}.
The appearance of the $\chi^{-2}$ scaling in $x$ represents the main qualitative deviation from the inverse participation ratio results in Eq.~\eqref{eq:IPRrandomMPS}.
Nevertheless, this modification does not affect the overall conclusion: a bond dimension scaling as $\chi = O(\mathrm{poly}(N))$ is sufficient for the ensemble to converge to the OCPT distribution.

Until now, the discussion has focused on real Clifford circuits. 
To go beyond this regime, we investigate the impact of non-Clifford (\textit{magic}) and non-real (\textit{imaginary}) resources (and the interplay thereof) on anticoncentration and state design properties.
We quantify these resources via established measures of magic~\cite{liu2022manybody, Heinrich2019robustnessofmagic, turkeshi2025pauli, leone2022stabilizer, piroli2021quantum, tirrito2024quantifyingnon} and imaginarity~\cite{wu_immresource2021, wu_operational_2021, wu_resource_2024, zhang_easily_2025, li_multi-state_2025, fernandes_unitary-invariant_2024}.

We focus on the doped $\mathbb{R}\mathrm{MPS}$; similar conclusion apply to glued real Clifford circuits.
We inject resources via the initial state, taking $|\Psi_{(r)}\rangle = |\psi\rangle^{\otimes r} \otimes |0\rangle^{\otimes N-r}$, where $|\psi\rangle$ represents the resource qubit.
We analyze three distinct directions of doping: real-magic $|H\rangle = \cos(\pi/8) |0\rangle + \sin(\pi/8)|1\rangle$, complex-magic $|T\rangle = \cos(\pi/8) |0\rangle + i\sin(\pi/8)|1\rangle$, and pure-imaginarity $|\pm i\rangle =(|0\rangle \pm i|1\rangle)/\sqrt{2}$.
Essentially, the doping modifies the boundary conditions of Eq.~\eqref{eq:IPRrandomMPS}, leading to a result dependent on the stabilizer purities of the chosen resource state. In the End Matter we show explicitly the development of this computation, while here we focus on the results. 
First, we consider doping with real magic states $|H\rangle$. Here, the modified boundary condition tends to privilege stabilizer purities of elements in the orthogonal commutant.  Thus, recalling the moments of the orthogonal ensemble $I_k^{\mathbb{R}\mathrm{Haar}}=(2k-1)!!/{\prod_{m=1}^{k-1}{(2^N+2m)}}$~\cite{sauliere2025universalityanticoncentrationchaoticquantum}, we obtain the asymptotic behavior 
\begin{equation}
    I_k^\mathrm{d\mathbb{R}\mathrm{MPS}}(r) \simeq_{N\to\infty} I_k^{\mathbb{R}\mathrm{Haar}} e^{{(2^{k-1}-1)}x} \;,
    \label{eq:dopedIPROrtho}
\end{equation}
where $x \simeq N/\chi$. Thus, doping with $r = \log_2(\chi)= O(\log N)$ real-magic states saturates the distribution to the orthogonal prediction.

Analogously, for complex magic states $|T\rangle$, the surviving terms are given by the symmetric group $S_k \subset \Xi_k$, which spans the unitary commutant.
The sum over purities converges to $|S_k| = k!$, the dimension of the unitary commutant.
Matching this with the unitary Haar moments $I_k^{\mathbb{C}\mathrm{Haar}}=k!/{\prod_{m=1}^{k-1}{(2^N+m)}}$, we find
\begin{equation}
    I_k^\mathrm{d\mathbb{R}\mathrm{MPS}}(r) \simeq_{N\to\infty} I_k^{\mathbb{C}\mathrm{Haar}} e^{\frac{2^k-2}{2}x} \;,
    \label{eq:dopedIPRunitary}
\end{equation}
implying that logarithmic complex-magic doping suffices to drive the system toward Porter-Thomas statistics.

Finally, we consider doping with imaginary stabilizer states $|\pm i\rangle$.
Remarkably, a single imaginary resource ($r=1$) is sufficient to break the real symmetry, as we prove in \cite{SM}. All operators in the complex Clifford commutant basis $\Sigma_k$ satisfy $\zeta_\pi(|\pm i\rangle) = 1$, yielding the sum $\sum_{\sigma \in \Xi_k} \zeta_\sigma = (-1;2)_{k-1}$.
This immediately recovers the Complex Clifford moments~\cite{magni2025anticoncentrationcliffordcircuitsbeyond}
\begin{equation}
    I_k^\mathrm{d\mathbb{R}\mathrm{MPS}}(1) \simeq I_k^{\mathbb{C}\mathrm{Stab}} e^{\frac{2^k-2}{2^2}x} \;.
    \label{eq:dopedIPRClifford}
\end{equation}
This result aligns with resource-theoretic findings that real operations plus a single imaginary state generate the full complex basis~\cite{wu_immresource2021, hickey_quantifying_2018, zhang_can_2024}.
Our results imply that a doping rate $r \sim O(\log^{1+c} N)$ (with $c>0$) suppresses deviations faster than any polynomial.

Extending this analysis to the frame potential, we find that the design properties follow the hierarchy established by the IPRs. Doping with polylogarithmic resources yields the asymptotic convergence:
\begin{equation}
    \mathcal{F}_k^{\mathrm{doped}} \to \begin{cases}
    \mathcal{F}_k^{\mathbb{R}\mathrm{Haar}} & \text{if } |\Psi_{(r)}\rangle = |H\rangle^{\otimes r}\otimes |0\rangle^{\otimes N-r} \\
    \mathcal{F}_k^{\mathbb{C}\mathrm{Haar}} & \text{if } |\Psi_{(r)}\rangle = |T\rangle^{\otimes r}\otimes |0\rangle^{\otimes N-r} \\
    \mathcal{F}_k^{\mathbb{C}\mathrm{Stab}} & \text{if } |\Psi_{(1)}\rangle = |\pm i\rangle\otimes |0\rangle^{\otimes N-1}
    \end{cases} \;.
\end{equation}

\section{Discussion} 
In this work, we have provided a comprehensive characterization of anticoncentration and state design in real Clifford tensor networks and circuits. By establishing the real Clifford ensemble as a minimal baseline, we disentangled the distinct roles of magic and imaginarity in driving the emergence of quantum chaos. We summarized these findings in Tab.~\ref{tab:results}, which maps a precise hierarchy of statistical regimes connecting real-Clifford, real-magic, and complex-magic ensembles. 
Crucially, our analysis reveals a sharp separation of scales in resource requirements: while approaching Haar randomness necessitates a polylogarithmic doping of magic states, recovering full unitary Clifford statistics requires only a \textit{single} imaginary resource. This observation suggests a fine-grained strategy for resource management in quantum architectures: a unitary state design can be efficiently constructed by doping a real Clifford backbone with real magic states and a solitary imaginary qubit, obviating the need for full complex-magic control. 
These results have immediate implications for quantum computation based on rebits~\cite{mckague2013power,koh2018quantum, delfosse2015wigner, rudolph20022rebitgateuniversal, tournaire20243dlatticedefectefficient} and inform protocols for real randomized benchmarking~\cite{Hashagen2018} by quantifying the exact resource costs to make those scalable in practice \cite{helsen_general_2022,heinrich2023randomizedbenchmarkingrandomquantum}. Looking forward, a natural extension of this framework is the inclusion of noise models~\cite{lovaz2024,turkeshi2024errorresilience,sauliere2025universalityanticoncentrationnoisyquantum} and their impact on applications like classical shadows \cite{helsen_thrifty_2023,chen_nonstabilizerness_2024,brieger_stability_2025}. Understanding how decoherence competes with the injection of magic and imaginarity will be fundamental to establishing the practical bounds of these statistical transitions in near-term devices.

\begin{acknowledgments}
\textbf{Acknowledgments.}
This work is dedicated to Ada. 
B.M., M.H., and X.T. acknowledge support from DFG under Germany's Excellence Strategy – Cluster of Excellence Matter and Light for Quantum Computing (ML4Q) EXC 2004/1 – 390534769, and DFG Collaborative Research Center (CRC) 183 Project No. 277101999 - project B01.

\textbf{Data and Code Availability}
Data and Code will be publicly shared at publication.
\end{acknowledgments}

\bibliography{biblio}

\clearpage

\begin{center}
    \textbf{End Matter}
\end{center}

\section{Review on anticoncentration of unitary, orthogonal and Clifford ensemble}
We collect here the key facts about anticoncentration in Haar-random unitary, orthogonal and (complex) Clifford circuits, which serve as the benchmark for our results on doped real Clifford circuits.

A Haar-random state on $N$ qubits, of local dimension $d$ is obtained by applying a Haar-distributed unitary $U$ to a reference computational-basis vector $|\pmb{0}\rangle$, yielding $|\Psi\rangle = U|\pmb{0}\rangle$.  
As discussed in the Main Text, the rescaled overlap 
\(
w = d |\langle \pmb{0}|\Psi\rangle|^2
\)
is distributed according to the Porter-Thomas law,
\begin{equation}
\mathcal{P}_\mathcal{U}(w) = \frac{d-1}{d}\left(1-\frac{w}{d}\right)^{d-2}
\;\xrightarrow[N\gg 1]{}\; e^{-w}.
\end{equation}

Similarly, for real Haar-random states, $\ket{\Phi} = O\ket{\pmb{0}}$ where $O \in \mathcal{O}(2^N)$, the distribution of the overlaps is a chi-squared distribution~\cite{sauliere2025universalityanticoncentrationchaoticquantum}, 
\begin{align}
    \mathcal{P}_\mathcal{O}(w)&= \frac{\Gamma(d/2)}{\sqrt{d}\Gamma((d-1)/2)}\frac{1}{\sqrt{\pi w}}(1-\frac{w}{d})^{(d-3)/2}\,\nonumber \\ &\xrightarrow[N\gg 1]{} \frac{1}{\sqrt{2\pi w}}e^{-\frac{w}{2}}
\end{align}

These results can be extrapolated from the computation of the inverse participation ratios, which are the moments of the distribution,
\begin{align}
&I_k^{\mathbb{C}\mathrm{Haar}} = d^{(1-k)}\,\mathbb{E}[w^k] = \frac{k!}{\prod_{m=0}^{k-1}(d+m)} \label{eq:haarvalU}\\  
&I_k^{\mathbb{R}\mathrm{Haar}} = \frac{(2k-1)!!}{\prod_{m=0}^{k-1}(d+2m)}.\label{eq:haarvalO}  
\end{align}
where $k!$ and and $(2k-1)!!$ equals the size of the commutant of the corresponding ensemble. The commutant is generated by the permutation group $\mathcal{S}_k$ for the unitary group and the pairing group over $2k$ elements, $\Theta_{k}$, for the orthogonal one. 
The orthogonal commutant, also known as the Brauer algebra~\cite{collins2006integration}, is isomorphic to the permutation enhanced by the partial transposition over certain replicas~\cite{zhang2006permutationpartialtranspose}. Given an operator over $k$ replicas, $A = a_1\otimes ... \otimes a_k$, the partial transposition over over the $i$th replica is defined as $A^{T_i} = a_1 \otimes a_2 \otimes ... \otimes a^T_i \otimes ... \otimes a_k $.
A representative element of this kind is the crossing operator
\begin{equation}
    \mathcal{X} = \frac{1}{d}\sum_{i,j =1}^{d-1}\ket{ii}\bra{jj} = \frac{1}{d}\sum_{P\in\mathcal{P}_1} \xi(P)P^{\otimes 2}
\end{equation}
which is the $\mathrm{SWAP}$ operator with partial transposition performed over the second replica. In terms of the Pauli matrices, the operator is multiplied by  $\xi(P) = \mathrm{Tr}(PP^T)$.

In physically local chaotic circuits, anticoncentration emerges only after a finite depth. Random unitary and orthogonal matrix product states reach the Haar value once their bond dimension grows polynomially with system size~\cite{lami2025anticoncentration}, while local quantum circuits converge at depths that scale logarithmically in the number of qubits~\cite{dalzell2022random,Christopoulos_2025}.  
In both scenarios, the inverse participation ratios can be written as the Haar predictions~(\ref{eq:haarvalU}) multiplied by a log-normal correction,
\begin{equation}
I_k^{\mathcal{U}} = I_k^{\mathbb{C}\mathrm{Haar}}\, e^{\frac{k(k-1)}{2}x}\quad \text{and} \quad I_k^{\mathcal{O}} = I_k^{\mathbb{R}\mathrm{Haar}}\, e^{k(k-1)x}\,.
\end{equation}
The full overlap distribution for general random circuits was also obtained in~\cite{lami2025anticoncentration,Christopoulos_2025}, where the parameter $x$ effectively plays the role of a fitting coefficient.

We now recall the results for complex Clifford circuits obtained in Ref.~\cite{magni2025quantumcomplexitychaosmanyqudit}. Given a random stabilizer state $\ket{\psi} = C\ket{\boldsymbol{0}}$ with $C\in \mathcal{C}_N$, the overlap distribution reads 
\begin{equation}
    \mathcal{P}(g) = \binom{N}{g}_2 \frac{2^{g(g+1)/2}}{(-2; 2)_N}\,,
\end{equation}
with the variable $g$ being the participation entropy of the state. From this distribution, named Clifford-Porter-Thomas (CPT), we can compute the inverse participation ratios
\begin{equation}
    I_k^{\mathbb{C}\mathrm{Stab}} = \frac{d^{(1-k)}(-1;2)_{k-1}}{(-d^{-1};2)_{k-1}}\,,
    \label{eq:stab-ipr}
\end{equation}
where ${(a;\xi)_n \equiv \prod_{m=0}^{n-1}(1-a\xi^m)}$ is the q-Pochhammer symbol~\cite{Kac2002}.
As before, the expression is obtained by unraveling the average over the Clifford group. In this case, the commutant is indexed by the set of stochastic Lagrangian subspaces of $\mathbb{Z}_2^{2k}$. These subspaces $\sigma$ satisfy the following conditions: (i) for every $(x,y)\in \sigma$ it holds $\sum_{i=0}^{k-1} (x_i^2 - y_i^2) = 0 \,\mathrm{mod}\,D$, where $D=4$ for qubits, (ii) $\sigma$ has dimension $k$ as a vector space, (iii) the vector $\pmb{1}_{2k}=(1,1,\dots,1)$ is an element of $\sigma$. 
We denote the set of all such stochastic Lagrangian subspaces as $\Sigma_k$, which has cardinality
\begin{equation}
|\Sigma_k| = \prod_{m=0}^{k-2}(2^m+1)     = (-1;2)_{k-1}\,.
\label{eq:sizesigma}
\end{equation}  
For any given stochastic Lagrangian subspace $\sigma$ we associate the corresponding vector $|\sigma\rrangle =\sum_{(x,y)\in \sigma}|x,y\rrangle$ on an individual replica qubit $\mathcal{H}^{\otimes 2k}$. 
The vectors $|\sigma\rrangle$ are linearly independent when $N \geq k-1$; in that case, $|\mathrm{Comm}_k(\mathcal{C}_{N})| = |\Sigma_k|$.
One of the operators inside the commutant, for $k=4$, is $\Omega_4 = \sum_{P\in \mathcal{P_1}} {(P\otimes P^\dagger)^{\otimes 2}}/{2}$ which remarkably corresponds to the projector of the stabilizer Rényi entropy~\cite{leone2022stabilizer}. 
Through the use of the property of the commutant and tensor network methods, various results are obtained, as shown in Tab.~\ref{tab:results}, for tensor networks and local circuits, also when doping with magic states.

\section{Doped real Clifford RMPS}
Here we provide a detailed derivation of the results for the doping with real and complex magic of $\mathbb{R}MPS$. 

The doping modifies the boundary condition of the $k$-replica circuit, which leads to 
\begin{equation}
\begin{split}
    I_k^{\mathrm{d}\mathbb{R}\mathrm{MPS}} &= \sum_{\{\pi_j,\sigma_j\}} \prod_{j=1}^{N-r-1} \left[ \mathrm{Wg}_{\pi_j,\sigma_j}(2\chi)G_{\sigma_j,\pi_{j+1}}(\chi) \right] \\
    &\quad \times \mathrm{Wg}_{\pi_{N-r},\sigma_{N-r}}(2\chi) \llangle \Psi_{(r)} | \pi_1 \rrangle \llangle \sigma_{N-r} | \mathbf{0} \rrangle \;.
\end{split}\label{eq:doped_path_integral}
\end{equation}
Iteratively resumming the bulk yields the same transfer matrix structure as the undoped case. However, the boundary term now contributes a factor $\zeta_{\pi_1}(|\Psi_{(r)}\rangle)$. Exploiting the product structure of the resource state, this factorizes as $\zeta_{\pi_1}(|\Psi_{(r)}\rangle) = [\zeta_{\pi_1}(|\psi\rangle)]^r$, where $\zeta_{\sigma}(|\psi\rangle) = \langle \psi|^{\otimes k} \sigma |\psi\rangle^{\otimes k}$ defines the real stabilizer purity of the state.
The exact IPR for the doped ensemble is thus 
\begin{equation}
    I_k^\mathrm{d\mathbb{R}\mathrm{MPS}}(r) = 2^{(1-k)N} \frac{\left[(-2^{-r+1};2)_{k-1}\right]^{N-r-1}}{\left[(-2^{-r};2)_{k-1}\right]^{N-r}}
    \left(\sum_{\sigma\in \Xi_k} \zeta_\sigma^r \right) \;.
    \label{eq:dopedIPRgeneric}
\end{equation}
The physics of the doping is entirely encoded in the specific values of the stabilizer purities $\zeta_\sigma$ for the chosen resource state $|\psi\rangle$.
As an example, we explicitly develop the computation for the real magic state $\ket{H}$.
We note that the basis $\Xi_k$ contains a subset $\Theta_k \subset \Xi_k$ that spans the commutant of the orthogonal group, with $|\Theta_k|=(2k-1)!!$.
The real stabilizer purity of the resource state gives $\zeta_\pi(|H\rangle) = 1$ for $\pi \in \Theta_k$, while $\zeta_\pi(|H\rangle) < 1$ otherwise, see~\cite{SM} for a proof. 
Introducing the generalized entropies \cite{turkeshi2023measuring}
\begin{equation}
    \mathcal{M}_\sigma = -\log_2\zeta_\sigma\,,
    \label{eq:gen_entropies}
\end{equation}
the counting becomes 
\begin{equation}
\begin{split}
    \sum_{\sigma\in \Xi_k}\zeta_\sigma^r &= (2k-1)!! + \sum_{\sigma \in \Xi_k/\Theta_k}\zeta_\sigma^r \nonumber \\ &= (2k-1)!! + \sum_{\sigma \in \Xi_k/\Theta_k}\left(\frac{x_0}{N}\right)^\mathcal{M_\sigma}\,,
\end{split}
\end{equation}
where, in the last step, we have used that $r = \log_2(\chi) = \log_2(N/x_0)$ and Eq.~\eqref{eq:gen_entropies}. Thus, for $x_0$ constant and $N \rightarrow \infty$, using the definition of the orthogonal inverse participation ration in Eq.~\eqref{eq:haarvalO}, we obtain
\begin{equation}
    I_k^\mathrm{d\mathbb{R}\mathrm{MPS}}(r) \simeq_{N\to\infty} I_k^{\mathbb{R}\mathrm{Haar}} e^{{(2^{k-1}-1)}x} \;,
\end{equation}
where $x \simeq N/\chi$, which is the result presented in the Main Text. 
\end{document}


\title{Supplemental material for Anticoncentration and State Design of Doped Real Clifford Circuits and Tensor Networks}

\author{Beatrice Magni~\orcidlink{0009-0009-8577-0525}}
\email{bmagni@uni-koeln.de}
\affiliation{Institut für Theoretische Physik, Universität zu Köln, Zülpicher Strasse 77, 50937 Köln, Germany}

\author{Markus Heinrich~\orcidlink{0000-0002-1334-7885}}
\affiliation{Institut für Theoretische Physik, Universität zu Köln, Zülpicher Strasse 77, 50937 Köln, Germany}

\author{Lorenzo Leone~\orcidlink{0000-0002-0334-7419}}
\affiliation{Dipartimento di Ingegneria Industriale, Università degli Studi di Salerno,
Via Giovanni Paolo II, 132, 84084 Fisciano (SA), Italy}
\affiliation{Dahlem Center for Complex Quantum Systems,
Freie Universität Berlin, 14195 Berlin, Germany}

\author{Xhek Turkeshi~\orcidlink{0000-0003-1093-3771}}
\email{xturkesh@uni-koeln.de}
\affiliation{Institut für Theoretische Physik, Universität zu Köln, Zülpicher Strasse 77, 50937 Köln, Germany}

\date{\today}

\maketitle

\section{Review on stabilizer tableau}
Here, we briefly review the characteristic formalism of stabilizer states. Given a system of $N$ qubits, we can define the set of Pauli strings as
\begin{equation}
    \mathcal{P}_{N} = \{\mathbb{1},X,Y,Z\}^{\otimes N} = \{i^{a\cdot b} X_1^{a_1}Z_1^{b_1}X_2^{a_2}Z_2^{b_2} \dots X_N^{a_N}Z_N^{b_N} | a_i,b_i \in \mathbb{Z}_2\} \,,
    \label{eq:paulistring}
\end{equation}
where the phase comes from the relation $XZ=-i Y$ and the exponent $a\cdot b = \sum_{i=1}^N a_i b_i$ is meant to be evaluated modulo four.
The Clifford group $\mathcal{C}_N$ is the subgroup of the unitary group $\mathcal{U}(2^N)$ generated by the Hadamard gate \(H = 1/\sqrt{2}\sum_{m,n=0}^1 (-1)^{mn} \ket{m}\bra{n} \), the controlled-NOT gate \(\text{CNOT} = \sum_{m,n=0}^1  \ket{m, m \oplus n}\bra{m, n}\), and the phase gate \(S = \sqrt{Z}\). 
Crucially, it maps Pauli strings to other Pauli strings up to a phase:
\begin{equation}
    \forall C\in\mathcal{C}_N \text{ and } P\in\mathcal{P}_N: \; CPC^\dagger \in \pm \mathcal{P}_N \,  \,.
    \label{eq:cliffgroup}
\end{equation} 

Stabilizer states are defined by applying Clifford operators to $\ket{\pmb{0}} := \ket{0}^{\otimes N}$, and we denote the set of stabilizer states as $\mathrm{Stab}_N := \{C\ket{\pmb{0}}|C\in \mathcal{C}_N\}$. Due to the properties of the Clifford group, these states are joint eigenstates of $N$ independent and commuting Pauli strings, $P_j \ket{\Psi} = (-1)^{\varphi_j} \ket{\Psi}$ for $j = 1, ..., N$, which uniquely characterize them. Indeed, since $Z_j\ket{\pmb{0}} = \ket{\pmb{0}}$ for each $j$, the state $\ket{\Psi}$ is stabilized by $C Z_j C^\dagger =: (-1)^{\varphi_j} P_j$, which allow us to rewrite the state as $\ket{\Psi}\bra{\Psi} = \prod_{j=1}^N\frac{\mathbb{1} + (-1)^{\varphi_j} P_j}{2}$. 
Conveniently, the \emph{stabilizers} $(-1)^{\varphi_j} P_j$ can be expressed as rows of a $(2N+1) \times N$ matrix, know as the \emph{stabilizer tableau}, as $\mathbf{S} = (\varphi_j|a^{(j)}|b^{(j)})_{j=1,...,N}$, where the parameters $a^{(j)}, b^{(j)} \in \mathbb{Z}_2^N$ and $\varphi_j \in \mathbb{Z}_2$ fix $P_j$ unambiguously as in Eq.~\eqref{eq:paulistring}. 
Remarkably, the stabilizer tableau is invariant under gaussian elimination, and we can decompose it in two $N \times N$ submatrices, $\mathbf{X} = (a_i^{(j)})_{i,j = 1, ..., N}$ and $\mathbf{Z} = (b_i^{(j)})_{i,j = 1, ..., N}$, plus a column vector representing the phases. These considerations are central in the computation of the overlap distribution in the Main Text.
Finally, the action $\ket{\Psi '} = C \ket{\Psi}$ of a Clifford operation $C$ on the stabilizer state $\ket\Psi$ can be represented as a linear mapping on $(a^{(j)}, b^{(j)})\in\mathbb{Z}_2^{2N}$ and a quadratic mapping on the phases, meaning that the resulting tableau $\mathbf{S}'$ characterizing the state $\ket{\Psi '}$ can be efficiently computed in $\mathrm{poly}(N)$ time and space~\cite{gottesman1997stabilizer}.

\section{Overview on the Real Clifford commutant}
Here, we present a general characterization of the real Clifford commutant, building on previous work on the unitary Clifford commutant. The following provides only the information necessary for the reader to have a self-contained understanding of the main results of this paper; a more in-depth discussion of the real Clifford commutant can be found in Ref.~\cite{bittel2025realclifford}.

We note that a unitary matrix $U$ is orthogonal, $U^T=U^{-1}=U^\dagger$, iff it has only real entries.
We call the subgroup $\mathcal{R}_N = \mathcal{C}_N \cap \mathcal{O}(2^N)$ of orthogonal Clifford unitaries, or alternatively the ones with real entries, the \emph{real Clifford group}.
We can define a basis for its commutant $\Xi_k$ --referred to as the real Clifford commutant in the following-- as
\begin{equation}
    \Xi_k = \text{span}\{\Phi_\mathcal{R}^{(k)}(P): \, P \in \mathcal{P}_N^{\otimes k}\},
    \label{eq:comm_basis}
\end{equation}
where the twirling operator of a generic operator $O$ is defined as 
\begin{equation}
    \Phi_\mathcal{R}^{(k)}(O) = \frac{1}{|\mathcal{R}_N|}\sum_{C\in \mathcal{R}_N} C^{\otimes k } O C^{\dagger\otimes k}\,.
    \label{eq:twirling}
\end{equation}
Building on Ref.~\cite{bittel2025completetheorycliffordcommutant}, we recall that a Pauli operator $P\in\mathcal{P}_N^{\otimes k}$ admits a decomposition of the form $P = \phi P_1^{\otimes v_1}P_2^{\otimes v_2}...P_m^{\otimes v_m}$, given $P_1,..., P_m \in \mathcal{P}_N\setminus\{\mathbb 1\}$ a set of algebraically independent traceless Paulis, an integer $m \leq k $ referred to as the \textit{order}, $\{v_i\}_{i=1,...,m} \in\mathbb{Z}_2^k $ a set of linearly independent binary vectors, and $\phi = \{\pm 1, \pm i\}$ (cfr. Lemma 13 in Ref.~~\cite{bittel2025completetheorycliffordcommutant}). Since Pauli operators either commute or anticommute, we can associate to $P_1,\ldots,P_m$ a binary matrix $G \in \mathrm{Sym}_0(\mathbb{Z}_2^{m\times m})$\footnote{$\mathrm{Sym}_0(\mathbb{Z}_2^{m\times m})$ is the space of $m \times m $ symmetric matrices with entries in $\mathbb{Z}_2$ and zero diagonal}, the adjacency matrix of their anticommutation graph, where vertices corresponding to Pauli operators are connected if they anticommute. Crucially, the association between $P \in \mathcal{P}_N^{\otimes k}$ and the pair $(V,G)$, with $V \coloneq \{v_1,...,v_m\} \in \mathbb{Z}_2^{k\times m}$, is unique up to a gauge $A \in GL(\mathbb{Z}_2^{m\times m})$, which can be accounted defining the equivalence class $[V,G]$. Thus, we can define the map $\mathcal{M}: P \mapsto \mathcal{M}(P) = [V,G]$, which, together with the set $P_1,...,P_m$, uniquely identify $P$. 

As anticipated, in the following, we are interested in looking at the action of real Clifford operators on Pauli operators to characterize the twirling action of the formers. To this end, it is useful to endowe the Pauli operator $P$ with some more information. Given a decomposition $P\propto P_{1}^{\otimes v_1}\cdots P_m^{\otimes v_m}$, we can also associate a binary vector $\Gamma\in\mathbb{Z}_2^m$ such that $\Gamma_i=1$ iff the number of $Y$ Pauli forming the Pauli string $P_i\in\mathbb{P}_n$ is odd, where $\Gamma_i=0$ if this number is even. So now, we can associate $P$ to a triplet $(V,G,\Gamma)$. This association is unique up to a gauge $A\in GL(\mathbb{Z}_2^{m\times m})$, which transform $V,G,\Gamma$ accordingly, and, similarly to before, we can define a map $\mathcal{M}_{\mathcal{R}}\,:\, P\mapsto\mathcal{M}_{\mathcal{R}}(P)=[V,G,\Gamma]$ which maps a given Pauli $P\in\mathcal{P}_N^{\otimes k}$ to the associated equivalence class $[V,G,\Gamma]$ see Ref.~\cite{bittel2025realclifford}.

With this information, one is able to build an orthogonal graph-based basis for the real Clifford commutant, in analogy with the one in Ref.~\cite{bittel2025completetheorycliffordcommutant}.
\begin{theorem}[An orthogonal basis for the real Clifford commutant]
    The set of operators, 
    \begin{equation}
        \invOmega_I([V,G,\Gamma]) \coloneq \frac{1}{|\mathbb{S}|}\sum_{Q\in\mathbb{S}}\varphi(Q)Q
    \end{equation}
    where $\mathbb{S} = \{Q\in \mathcal{P}_N^{\otimes k} | \mathcal{M}_{\mathcal{R}}(Q) = [V,G,\Gamma] \} $, $\varphi(Q)\coloneq \tr(QT_{(k...21)})/2^N$ with $T_{(k...21)}$ cyclic permutation over $k$ replica, $V\in \mathrm{Even}(\mathbb{Z}_2^{k\times m})$, $G\in \mathrm{Sym_0}(\mathbb{Z}_2^{m\times m })$ and $\Gamma\in \mathbb{Z}_2^m$, form an orthogonal basis for the real Clifford commutant $\Xi_k$ for any $k$ and $N$.
\end{theorem}
\begin{proof}
The proof closely follows the proof of Theorem 25 of Ref.~\cite{bittel2025completetheorycliffordcommutant}.
    As stated before, we can construct the real Clifford commutant from the image of the twirling operator acting over Pauli operators (Eq.~\eqref{eq:comm_basis}).
    From Eq.\eqref{eq:twirling}, we have
    \begin{equation}
        \Phi_\mathcal{R}^{(k)}(P)=\frac{1}{|\mathcal{R}(2^N)|}\sum_{C\in \mathcal{R}(2^N)} C^{\otimes k } P C^{\dagger\otimes k} = \frac{1}{|\text{orb}_{\mathcal{R}}(P)|}\sum_{Q\in \text{orb}_{\mathcal{R}}(P) } Q\,,
    \end{equation}
    where $\text{orb}_{\mathcal{R}}(P) \coloneq \{Q\in \pm \mathcal{P}_N^{\otimes k} |\,\exists C\in \mathcal{R} : Q = C^{\otimes k } P C^{\dagger\otimes k} \}$ is the orbit of Pauli $P$ under the action of real Clifford operators.
    Since $C^{\otimes k}$ commutes with $T_{(k...21)}$, we have $\varphi(P)=\varphi(\Phi_\mathcal{R}^{(k)}(P))$ and hence $\Phi_\mathcal{R}^{(k)}(P) = 0$ implies $\varphi(P)= 0$.
    By the same line of reasoning as in Ref.~\cite{bittel2025completetheorycliffordcommutant}, the converse is also true and we repeat the argument in the following.
    Inserting the representation $P = \phi P_{1}^{\otimes v_1} \cdots P_{m}^{\otimes v_m}$ in the definition of $\varphi(P)$ yields
    \begin{equation}
        \varphi(P) = \frac{1}{2^N} \tr(P T_{(k\ldots21)})
        = \frac{1}{2^N}\, \phi \, \tr\!\left(
            \prod_{i=1}^{m} \prod_{\alpha=1}^{k} P_i^{(v_i)_\alpha}
        \right) \,.
    \end{equation}
    Note that $\varphi(P) = 0$ if and only if the product of Pauli operators in the trace is traceless.
    Using the binary representation $u^{(i)}:=(a^{(i)},b^{(i)})$ of the $P_i$ (cf.~Eq.~\eqref{eq:paulistring}).
    this means
    \begin{equation}
        0 \neq \sum_{i,j} (v_i)_j \, u^{(i)}
    = \sum_i |v_i| \, u^{(i)} \pmod 2 \,,
    \end{equation}
    where $|v_i| = \sum_j (v_i)_j$ the Hamming weight of the vector.
    Since the $P_i$ are traceless and algebraically independent, the $u^{(i)}$ are non-zero and linearly independent.
    Thus, $\varphi(P) = 0$ implies that $|v_i| = 1 \mod 2$ for some $i$.
    Now consider a Clifford that maps $P_i\mapsto -P_i$ and leaves the $P_j$ with $j\neq i$ invariant.
    Then, $P$ and $-P$ are in the same orbit and we conclude that $\Phi_\mathcal{R}^{(k)}(P)=\Phi_\mathcal{R}^{(k)}(-P)=-\Phi_\mathcal{R}^{(k)}(P)$, hence $\Phi_\mathcal{R}^{(k)}(P)=0$.

    Note that the above argument also shows that for every $P$ that is not annihilated under the twirl, the vectors $v_i$ in $P = \phi P_{1}^{\otimes v_1} \cdots P_{m}^{\otimes v_m}$ have to have even Hamming weight, i.e.~$V\in\mathrm{Even}(\mathbb{Z}_2^{k\times m})$.
    The action of a real Clifford operator on $P$ takes the form
    \begin{equation}
    C^{\otimes k} P C^{T \otimes k}
        = \phi\, C^{\otimes k} P_1^{\otimes v_1} \cdots P_m^{\otimes v_m} C^{T \otimes k}
        = \phi\, (CP_1C^T)^{\otimes v_1} \cdots (CP_mC^T)^{\otimes v_m} \,,
    \end{equation}
    since $C^\dagger = C^T$. Therefore, $V$, $G$ and also $\Gamma$ are preserved in the orbit; the latter follows from the fact that we are restricting the twirling to real Clifford opearations. Hence, $\mathcal{M}_{\mathcal{R}}(C^{\otimes k} P C^{\otimes k}) = \mathcal{M}_{\mathcal{R}}(P)$. Defining $\mathbb{S} = \{ Q \in \mathbb{P}_{n}^{\otimes k} \mid \mathcal{M}_{\mathcal{R}}(Q) = [V,G,\Gamma]  \}$, we can identify the twirling operator of $P$ as
    \begin{equation}
        \Phi_{\mathcal{R}}^{(k)}(P)
        = \frac{\varphi^{*}(P)}{|\mathbb{S}|}
          \sum_{Q \in \mathbb{S}} \varphi(Q)\, Q , 
          \label{eq:proof_basis}
    \end{equation}
    where the relative $\pm$ phase between Pauli operators in $\mathrm{orb}_\mathcal{R}(P)$ is given by $\varphi(P)/\varphi(Q)$.  
    Indeed, under Clifford conjugation
    $\phi(P)$ is invariant, meaning that, if $C^{\otimes k} P C^{\dagger \otimes k} = \theta Q$ with $\theta=\pm 1$,  we have 
   $ \phi(P) = \phi\!\left(C^{\otimes k} P C^{\dagger \otimes k}\right)
            = \theta \phi(Q).$
    This directly implies $\theta = \phi(P)/\phi(Q) = \varphi(Q)/\varphi(P) = \varphi^{*}(P)\varphi(Q),$
    which establishes Eq.~\eqref{eq:proof_basis}.  
    Hence, we may express the twirling operator as
    \begin{equation}
    \Phi_{\mathcal{R}}^{(k)}(P)
          = \varphi^{*}(P)\, \invOmega_{I}([V,G,\Gamma])
          \qquad\text{where}\qquad [V,G,\Gamma]=\mathcal{M}_{\mathcal{R}}(P).
     \end{equation}
    Thus, we have shown that the operators $\invOmega_{I}([V,G,\Gamma])$ generate the commutant. Moreover, they are mutually orthogonal  with respect to the Hilbert-Schmidt inner product since the orbits are disjoint, i.e.
    \begin{equation}
        \tr\left(\invOmega_{I}([V,G,\Gamma])^{\dagger}\,\invOmega_{I}([V',G',\Gamma'], )\right)
        = 0
        \qquad \text{for } [V,G,\Gamma] \neq [V',G',\Gamma']\,.
    \end{equation}

\end{proof}
As done in Ref.~\cite{bittel2025completetheorycliffordcommutant} and presented in Ref.~\cite{bittel2025realclifford}, we can Fourier transform the graph dependent basis to obtain the generalized Pauli monomials, previously shown in Ref.~\cite{bittel2025operationalinterpretationstabilizerentropy}:
\begin{equation}
    \Omega(V,M,\Gamma)=\frac{1}{2^{Nk}}\sum_{P_1,\ldots,P_k}P_{1}^{\otimes v_1}\cdots P_{k}^{\otimes v_k}\prod_{i<j}\chi(P_i,P_j)^{M_{ij}}\prod_i\xi(P_i)^{\Gamma_i}\,,
    \label{eq:paulimonomials}
\end{equation}
where $V\in \mathrm{Even}(\mathbb{Z}_2^{k\times m})$, $M\in \mathrm{Sym_0}(\mathbb{Z}_2^{m\times m })$, $\Gamma\in \mathbb{Z}_2^m$, $\chi(P_i,P_j) = \tr(P_iP_jP_i^\dagger P_j^\dagger)$ and $\xi(P_i) = \tr(P_iP_i^T)$. In the following, we will simplify the notation using letters as $\sigma$, to encode the triplets $(V,M,\Gamma)$, which characterize the operator.

We can now proceed to prove the expressions in the main text for the dimension of the commutant and the free sum over the Gram matrix.
\begin{theorem}[Dimension of the real Clifford commutant]
    For $N\geq k-1$, the dimension of the real Clifford commutant is, for any $k$,
    \begin{equation}
        |\Xi_k| = \prod_{m=0}^{k-2}(2^{m+1}+1).
    \end{equation}
    Moreover, defined the Gram matrix as $G_{\sigma\pi} = \tr(\Omega_\sigma^\dagger \Omega_\pi)$, where $\Omega_\sigma$ is a generalized Pauli monomial, the free sum over one of its index is a constant equal to
    \begin{equation}
        \sum_{\sigma \in \Xi_k}G_{\sigma\pi}(2^N) =  2^N\prod_{m=0}^{k-2}(2^{m+1}+2^N).
    \end{equation}
\end{theorem}
\begin{proof}
    Having constructed an orthogonal basis of the real Clifford commutant, its
    dimension follows from counting the number of equivalence classes $[V,G,\Gamma]$ that
    label its elements. Since we are mainly interested in the case $N \ge k-1$, below we only provide a counting for this case, even if following the calculation of Ref.~\cite{bittel2025completetheorycliffordcommutant} the generalization to every $N,k$ is straightforward.
    
    For fixed $m$, the number of possible $V$-labels is the Gaussian binomial
    $\binom{k-1}{m}_2$, which counts the number of binary vectors subspaces of
    $\mathbb{Z}_2^{k}$ generated by $m$ vectors of even Hamming weight.  
    The anticommutation graph is encoded by a symmetric matrix
    $G\in\mathrm{Sym}_0(\mathbb{Z}_2^{m\times m})$, giving $2^{m(m-1)/2}$
    choices, and the vector $\Gamma\in\mathbb{Z}_2^m$ contributes an additional
    factor $2^m$.  
    
    Summing over \(m=0,\ldots,k-1\) gives
    \begin{equation}
       |\Xi_k|
       = \sum_{m=0}^{k-1} \binom{k-1}{m}_2\, 2^{m(m+1)/2}
       = \prod_{m=0}^{k-2} (2^{m+1}+1),
    \end{equation}
    where the product form follows from the Gaussian binomial theorem. To generalize it to every $N$ and $k$ it is sufficient to note that not all graphs are allowed, and the dimension can be computed explicitly following Theorem 26 of Ref.~\cite{bittel2025completetheorycliffordcommutant}.
    
    Similarly, it holds for the Gram matrix. Once defined the generalized Pauli monomial, they are characterized by a similar set of invariants, whose counting is the same as before. Additionally, each element of the Gram matrix weights $2^{Nm}$, as shown in Ref.~\cite{bittel2025completetheorycliffordcommutant}, giving the final expression
    \begin{equation}
        \sum_{\sigma \in \Xi_k}G_{\sigma\pi}(2^N) =  \sum_{m=0}^{k-1}\binom{k-1}{m}_22^{m(m+1)/2}\,2^{m(N+1)} = 2^N\prod_{m=0}^{k-2}(2^{m+1}+2^N).
    \end{equation}
\end{proof}

Alternatively, a basis for the real Clifford commutant may be obtained using the language of \emph{stochastic Lagrangian subspaces} from Ref.~\cite{gross2021schurweyl}.
We call a subspace $T\subset\mathbb{Z}_2^{k}\times \mathbb{Z}_2^{k}$ \emph{real stochastic Lagrangian} iff (1) $\dim(T)=k$, (2) $x\cdot x' + y \cdot y' = 0$ for all $(x,y),(x',y')\in T$ ($T$ is self-orthogonal), and (3) $1_{2k}\in T$.
Note that the  difference to the stochastic Lagrangian subspaces in Ref.~\cite{gross2021schurweyl} is that the mod 4 isotropy condition, $x\cdot x + y\cdot y=0 \mod 4$, is dropped but self-orthogonality of $T$ (w.r.t.~the dot product) is kept.
We denote the set of real stochastic Lagrangians by $\Sigma_k^{\mathbb{R}}$.
For any $T\in\Sigma_k^{\mathbb{R}}$ we define an operator
\begin{equation}
    r(T) := \sum_{(x,y)\in T} \ket{x}\bra{y} \,.
\end{equation}

\begin{theorem}
    For $N\geq k-1$, the operators $R(T):=r(T)^{\otimes N}$ define a basis of the real Clifford commutant.
\end{theorem}

\begin{proof}
    As in the proof of Lemma 4.5 in Ref.~\cite{gross2021schurweyl}, we can check that the $R(T)$ lie in the real Clifford commutant by checking that they commute with the generators of the real Clifford group.
    The only difference to Ref.~\cite{gross2021schurweyl} is that we have to replace the phase gate by a Pauli $Z$ (thus the lack of mod 4 isotropy in the definition).
    The proof of linear independence for $N\geq k-1$ in Lemma 4.7 in Ref.~\cite{gross2021schurweyl} holds unchanged for $\Sigma_k^{\mathbb{R}}$.
    Next, we have to count the number of real stochastic Lagrangians, to which end we follow the proof of Theorem 4.10 in Ref.~\cite{gross2021schurweyl}.
    There, for every $T$, the diagonal subspace $T_\Delta =T\cap \{ (x,x)\,|\, x\in\mathbb{Z}_2^k\}$ is studied.
    For a given $T_\Delta$ (with $1_{2k}\in T_\Delta$), the ways to complete it to a self-orthogonal, $k$-dimensional subspace $T$ are given by the number of possible complements $T = T_\Delta \oplus T_N$.
    Given a basis $v_1,\dots,v_{k-l},w_1,\dots,w_l$ of $\mathbb{Z}_2^k$ such that $T_\Delta=\{(v,v)\,|\,v\in V\}$ with $V:=\mathrm{span}\{v_1,\dots,v_{k-l}\}$, it is shown in Ref.~\cite{gross2021schurweyl} that each complement $T_N$ has a unique basis of the form $(\hat w_i +u_i, u_i)$ ($i=1,\dots,l$) where $\hat w_i$ are dual basis vectors ($\hat w_i\cdot w_j=\delta_{ij}$ and $\hat w_i \cdot v_j = 0$), and $u_i=\sum_{j=1}^k A_{ij}w_j$.
    Self-orthogonality then imposes the constraint 
    \begin{equation}
        0 = (\hat w_i +u_i, u_i)\cdot (\hat w_j +u_j, u_j) = \hat w_i \cdot \hat w_j + A_{ij} + A_{ji} \,.
    \end{equation}
    Note that this equation holds modulo 2 and implies that the lower triangular part of $A$ is determined by its upper triangular part.
    From this point on, the calculation deviates from Ref.~\cite{gross2021schurweyl}:
    Since we do not have the mod 4 isotropy condition, but only the mod 2 condition, we do not obtain constraints on the diagonal entries of $A$.
    This means that instead of $2^{l(l-1)/2}$, we now have $2^{l(l+1)/2}$ choices for $A$.
    The ways to choose a $(k-l)$-dimensional subspace $V$ are given by the number of $(k-l-1)$-dimensional subspaces in $\mathbb{Z}_2^k/\langle 1_k\rangle$, thus there are $\binom{k-1}{k-l-1}_2 = \binom{k-1}{l}_2$ many choices.
    Finally, we find
    \begin{equation}
        |\Sigma_k^{\mathbb{R}}| = \sum_{l=0}^{k-1} 2^{l(l-1)/2} 2^l \binom{k-1}{l}_2 = \prod_{m=0}^{k-2} (2^{m+1}+1) \,,
    \end{equation}
    by an application of the Gaussian binomial theorem.
\end{proof}

\section{Additional proofs}

\subsection{Proof for the maximal imaginary state}
As seen in the Main Text, the injection of a single $\ket{\pm i}$ in the real Clifford circuits drives the state design and anticoncentration exactly to that of unitary Cliffords. Here, we show why this is valid directly from the commutant perspective.

Let us show that $\tr(\Omega \sigma^{\otimes k})=0$ if $\sigma=\ket{+i}\bra{+i}$ and $\Omega$ belongs to the commutant of the real Clifford but not in the commutant of the Clifford group. From Eq.~\eqref{eq:paulimonomials}, which define the generalized Pauli monomial, the condition to belong to the real Clifford commutant and not to the Clifford commutant is simply that $\Gamma\neq0$. We have
\begin{align}
    tr(\Omega(V,M,\Gamma)\sigma^{\otimes k})=\frac{1}{d^k}\sum_{P_1,\ldots, P_k\in\mathbb{Y}}\tr(P_1^{\otimes v_1}\cdots P_{k}^{\otimes v_k}\sigma^{\otimes k})\prod_{i}\xi(P_i)^{\Gamma_i}
\end{align}
where $\mathbb{Y}$ is the Abelian group formed by $Y$-strings only. We restricted the sum only to that because $\ket{+i}\bra{+i}=\frac{\mathbb{I}+Y}{2}$ and nothing else can contribute to the sum, being $\mathbb{Y}$ a group. We also used that the group is Abelian and therefore $\chi(P_i,P_j)=0$ for $P_i,P_j\in\mathbb{Y}$. Now, if $\Omega$ does belongs to the real Clifford commutant and not to the Clifford commutant $\Gamma\neq 0$, hence there exists at least $j\in[m]$ such that $\Gamma_j\neq 0$. Without loss of generality, let us set $j=1$. We just need to note that we can write
\begin{align}
    \frac{1}{d^k}\sum_{P_1,\ldots, P_k\in\mathbb{Y}}tr(P_1^{\otimes v_1}\cdots P_{k}^{\otimes v_k}\sigma^{\otimes k})\prod_{i}\xi(P_i)^{\Gamma_i}=\frac{1}{2d^{k-1}}\sum_{P_2,\ldots,P_k}tr(P_2^{\otimes v_2}\cdots P_{k}^{\otimes v_k}\sigma^{\otimes k})-\frac{1}{2d^{k-1}}\sum_{P_2,\ldots,P_k}tr(P_2^{\otimes v_2}\cdots P_{k}^{\otimes v_k}\sigma^{\otimes k})=0
\end{align}
using that 1) $P_1^{\otimes v_1}\ket{\sigma}=\ket{\sigma}$ 2) that $\xi(P)=(-1)^{\#Y}$ and 3) in the group $\mathbb{Y}$ there are $d/2$ many Pauli strings with odd number of $Y$ and $d/2$ with even number of $Y$.

\subsection{Stabilizer purity}
In this subsection, we show that for the state $\ket{H}$ defined in the main text, we have that any generalized stabilizer purity but the one corresponding to expectation values of elements in the commutant of the orthogonal group are strictly less than $1$. Let $\Omega$ be a generalized Pauli monomial. Since $\Omega$ factorizes on qubits, we can consider directly $\tr(\Omega \ket{H}\bra{H}^{\otimes k})$. If $\Omega\in \Theta_k$ (i.e. the commutant of the orthogonal group), since $\ket{H}$ is real in the computational basis, it is immediate to see that $\tr(\Omega \ket{H}\bra{H}^{\otimes k})=1$. Let now $\Omega\in\Xi_k\setminus\Theta_k$. We discriminate two cases. A) $\Omega$ belongs to the commutant of the Clifford group. In this case, since $\ket{H}$ is not a stabilizer state, we know that $P_{4}(\ket{H}) := \tr(\Omega_4 \ket{H}\bra{H}^{\otimes k}) <1$ where $\Omega_4 = d^{-1}\sum_{P\in\mathcal{P}_N}P^{\otimes 4}$~\cite{leone2022stabilizer}. 
Moreover, in Ref.~\cite{bittel2025operationalinterpretationstabilizerentropy} it has been shown that $\tr(\Omega\ket{\psi}\bra{\psi}^{\otimes k})\le P_{4}(\ket{\psi})$ for any state $\ket{\psi}$. Putting these two results together, we have $\tr(\Omega\ket{H}\bra{H}^{\otimes k})<1$ as desired. B) $\Omega$ belongs to the commutant of the orthogonal Clifford but not in the commutant of the Clifford group. In the previous section, we noticed that for any such $\Omega$ there exists a partial transposition $T$ such that $\Omega^T$ belongs to the commutant of the Clifford group. We now use two facts. First, we use that the trace is invariant by partial transposition. Second, we use that $\ket{H}$ is real in the computational basis and therefore its partial transpose is equal to itself. We therefore have
\begin{align}
    \tr(\Omega\ket{H}\bra{H}^{\otimes k})=\tr(\Omega^T\ket{H}\bra{H}^{\otimes k})\le P_4(\ket{H})\le 1
\end{align}
which concludes the proof. 

The general proof to show that $\tr(\Omega\ket{\psi}\bra{\psi}^{\otimes k}) \leq 1$ for any (possibly not real) state can be done iteratively as the proof of Theorem 10 in Ref.~\cite{bittel2025operationalinterpretationstabilizerentropy}. Indeed, requiring that the span of the matrix $V$ does not contain weight $2$ elements we are essentially ruling out all the elements of the real orthogonal commutant (as well as the unitary commutant). Following exactly the same reasoning of the proof of Theorem 10, we can conclude that $\tr(\Omega\ket{\psi}\bra{\psi}^{\otimes k})\leq 1$ for any state $\ket{\psi}$. 
\color{black}

\bibliography{biblio}